\documentclass[12pt,eqsecnum]{revtex4}
\begin{document} 
\title{
  Relating Gribov-Zwanziger theory and Yang-Mills theory in Batalin-Vilkovisky formalism
}

\author{ Sudhaker Upadhyay\footnote {e-mail address: sudhakerupadhyay@gmail.com}}
\author{ Bhabani Prasad Mandal\footnote{e-mail address:
\ \ bhabani.mandal@gmail.com }}

\affiliation { Department of Physics,\\
Banaras Hindu University,\\
Varanasi-221005, INDIA. \\
}

\begin{abstract}
We consider the BRST invariant Gribov-Zwanziger theory with appropriate horizon term in 
Batalin-Vilkovisky formalism. The usual infinitesimal BRST transformation 
is generalized by considering the parameter finite and field dependent.
We show that such finite field dependent BRST transformation with suitable choice of finite parameter relates 
the generating functional of Gribov-Zwanziger theory to that of the Yang-Mills theory.
\end{abstract}

\maketitle 

\section{Introduction}
In non-Abelian gauge theories even after gauge fixing the redundancy of gauge 
fields is not completely removed in certain gauges for large gauge fields (Gribov ambiguity)
\cite{gri}. In order to resolve the Gribov ambiguity, Gribov and Zwanziger   
 restrict the domain of integration in the functional integral to the so-called Gribov region  $\Omega $ 
(defined in such a way 
that
the Faddeev-Popov (FP) operator is strictly positive. i.e.
 $ \Omega \equiv \{ A_\mu^a, \partial_\mu A_{\mu}^ a=0, {\cal{M}}^{ab} >0\} )$, whose 
boundary is known as  the first Gribov horizon \cite{zwan}. This has been
achieved by adding a nonlocal term, commonly known as horizon term, to the Yang-Mills (YM) action \cite
{zwan}. The non-Abelian gauge theories in those gauges contain so-called Gribov copies. 
Gribov copies play 
a crucial role in the infrared (IR) regime while it can be neglected in the perturbative 
ultraviolet (UV) regime \cite{gri,zwan}.  Such theories do not satisfy the 
 Kugo-Ojima (KO) criterion for color confinement \cite{ko} since the presence of the nonlocal horizon term  
breaks the  BRST invariance of the theory \cite{sore1}. 
Recently, it has been shown that the BRST symmetry of the Gribov-Zwanziger (GZ) theory can be restored by introducing 
the auxiliary fields  which explains the KO analysis\cite{sor}.

In the present work we generalize such extended BRST transformation for GZ theory by allowing the parameter to be 
finite and field 
dependent following the method developed in the reference  \cite{sdj}. Such a generalized BRST (FFBRST) 
transformation is also a symmetry of the   effective action.  However, being finite in nature such a transformation 
 does not leave the path integral measure as well as the generating functional invariant.
 We consider the  Batalin-Vilkovisky (BV) formulation of this theory.
For an appropriate choice of finite field dependent parameter we show that such 
a finite transformation relates the generating functional for GZ theory to the generating 
functional in YM theory in BV formulation. 
 
The paper is organized as follows. In Sec. II we discuss briefly the GZ theory and extended BRST symmetry.  Sec. 
III   is devoted to the construction of FFBRST transformation. Connection of GZ theory and YM theory 
in BV formalism is 
established in Sec. IV.
Last section is reserved for conclusions. 

\section{Preliminary: The GZ theory }
The generating functional for standard GZ theory \cite{zwan,sor, sb} reads as   
\begin{eqnarray}
Z_{GZ}=\int[ D\phi] e^{-S_{GZ}},\ \ S_{GZ} =S_{YM} +S_h,\label{zfun}
\end{eqnarray}
where  the YM effective action $S_{YM} $ in Landau gauge and local horizon term $S_h$  are written as
\begin{eqnarray}
S_{YM}&=&  \int d^4x\left[\frac{1}{4}F^a_{\mu \nu }F_{\mu \nu}^a + B^a\partial_\mu A_\mu ^a +\bar c^a\partial_\mu {
\cal D}_\mu 
^{ab}c^b\right],\nonumber\\
S_h &=& \int d^4x \left[\bar\varphi_i^{a}\partial_\mu{\cal D}_\mu^{ab}
\varphi^{b}_i-\bar\omega_i^{a}\partial_\mu{\cal D}_\mu^{ab}
\omega^{b}_i
- gf^{abc}\partial_\mu\bar\omega^a_i{\cal D}_\mu^{bd}c^d\varphi^c_i \right.\nonumber\\
&-&\left.\gamma^2 g\left( f^{abc} A^a_\mu\varphi^{bc}_\mu 
 +  f^{abc} A^a_\mu\bar\varphi^{bc}_\mu +\frac{4}{g}(N^2 -1)\gamma^2 
\right)\right].
\end{eqnarray}
The conventional BRST transformation for all the fields is given by
\begin{eqnarray}
\delta_b A_\mu^a &=& -{\cal D}_\mu^{ab}c^b\ \Lambda, \ \ \delta_b c^a =\frac
{1}{2}gf^{abc}c^bc^c
\ \Lambda, \ \
 \delta_b \bar c^a = B^a\ \Lambda, \ \delta_b B^a =0, \nonumber\\
 \delta_b\varphi_i^a &=& -\omega_i^a \ \Lambda,  \ \delta_b\omega_i^a =0,
\ \ \delta_b\bar\omega_i^a  = \bar \varphi_i^a \ \Lambda, \ \ \ \ \delta_b\varphi_i^a
 =0,\label{sym}
\end{eqnarray}
where $\Lambda$ is usual infinitesimal and global  BRST parameter and index $i (\equiv  \mu, a)$ is a composite index. 
Here we see that the BRST symmetry is broken softly for the GZ action \cite{zwan},
 the breaking is due to the presence of   $S_h$ term. 
To discuss the renormalizability of this theory the horizon term  $S_h$ is
 extended by introducing 3 doublets of sources 
$(U_\mu^{ai}, M_\mu^{ai}), (V_\mu^{ai}, N_\mu^{ai}) $  and $(T_\mu^{ai}, R_\mu^{ai})$, as
\begin{eqnarray}
\Sigma_\gamma &=&\int d^4 x \left [ -M^{ai}_\mu {\cal D}_\mu^{ab}\varphi^{bi} 
-gf^{abc}U^{ai}_\mu{\cal D}_\mu^{bd}c^d\varphi^{ci} +U^{ai}_\mu {\cal D}_\mu^{ab}
\omega^{bi} - N^{ai}_\mu {\cal D}_\mu^{ab}\bar\omega^{bi} \right.\nonumber\\
&-&\left.V^{ai}_\mu
{\cal D}_\mu^{ab}\bar\varphi^{bi} +gf^{abc}V^{ai}_\mu{\cal D}_\mu^{bd}c^d
\bar\omega^{ci}
- M^{ai}_\mu V^{ai}_\mu +U^{ai}_\mu N^{ai}_\mu -gf^{abc}R^{ai}_\mu{\cal D}_\mu^
{bd}c^d\bar\omega^{ci}\right.\nonumber\\
& +&\left.gf^{abc}T^{ai}_\mu{\cal D}_\mu^{bd}c^d\bar\varphi^{ci} \right ].
\end{eqnarray} 
The sources $M_\mu^{ai}, V_\mu^{ai}, R_\mu^{ai}$ are commuting and the remaining three
$U_\mu^{ai}, N_\mu^{ai}, T_\mu^{ai}$ are fermionic in nature.
The BRST symmetry transformation for this extended sources are given as
\begin{eqnarray}
\delta_b U^{ai}_\mu  = M_\mu^{ai}  \Lambda,  
\delta_b V^{ai}_\mu  = -N_\mu^{ai} \Lambda, 
\delta_b T^{ai}_\mu =-R_\mu^{ai} \Lambda, \delta_b M^{ai}_\mu =  \delta_b N^{ai}_\mu =  \delta_b R^{ai}_\mu =0.\label{syma}
\end{eqnarray}
To make the extended theory reminiscent with original theory (\ref{zfun}), 
at the end,   sources are equated to have following physical values \cite{sor}:
\begin{eqnarray}
U^{ai}_\mu |_{phys} =N^{ai}_\mu |_{phys}=T^{ai}_\mu |_{phys}=0,\  
M^{ab}_{\mu\nu} |_{phys} = V^{ab}_{\mu\nu} |_{phys}=R^{ab}_{\mu\nu} |_{phys}=\gamma^2
\delta^{ab}\delta_{\mu\nu},
\end{eqnarray}
such that $\Sigma_\gamma|_{phys}=S_h $.  
\section{Construction of FFBRST transformation in Euclidean space}
The properties of BRST transformation  do  not depend on  whether
the   parameter $\Lambda$    is (i) finite or infinitesimal, (ii) field dependent or not, 
as long as it is anticommuting in nature. Keeping this in mind, Joglekar and 
Mandal   introduced finite field dependent BRST transformation \cite {sdj}, which has 
found many applications in gauge field theories \cite{sdj,sb,   rb, ssb, susk, subm}.
Following these observations we 
make  the parameter $\Lambda$ given in Eqs. (\ref{sym}) and (\ref{syma}) 
  finite and field 
dependent without losing its properties. To 
 construct the FFBRST transformations we start 
with making the infinitesimal parameter field dependent by introducing a parameter $\kappa\ 
(0\leq \kappa\leq 1)$ and making all the fields, $\phi(x,\kappa)$, $\kappa$ dependent such 
that $\phi(x,\kappa =0)=\phi(x)$ and $\phi(x,\kappa 
=1)=\phi^\prime(x)$, the transformed field. 

The usual infinitesimal BRST transformations, thus can be written generically as 
\begin{equation}
{d\phi(x,\kappa)}=\delta_b [\phi (x,\kappa ) ]\Theta^\prime [\phi (x,\kappa ) ]{d\kappa},
\label{diff}
\end{equation}
where the $\Theta^\prime [\phi (x,\kappa ) ]{d\kappa}$ is the infinitesimal but field 
dependent parameter.
The generalized BRST transformations with the finite field dependent parameter then can be 
constructed by integrating such infinitesimal transformations from $\kappa =0$ to $\kappa= 1$, to obtain
\begin{equation}
\phi^\prime\equiv \phi (x,\kappa =1)=\phi(x,\kappa=0)+\delta_b[\phi(x) ]\Theta[\phi(x) ],
\label{kdep}
\end{equation}
where 
$\Theta[\phi(x)]=\int_0^1 d\kappa^\prime\Theta^\prime [\phi(x,\kappa^\prime)],  
$ is the finite field dependent parameter. Following this method, the  BRST 
transformation, in Eqs. (\ref{sym}) and (\ref{syma}), is generalized such that the parameter is finite and 
field dependent. 
Such generalized BRST transformation   
is symmetry  of the effective action. However, the  
path integral measure $[D\phi]$ in Eq. (\ref{zfun}) is not invariant under such 
transformation  as the 
BRST parameter is finite and field dependent. 
The Jacobian of the path integral measure (defined through Eq. $D\phi' =J(\kappa) D\phi
$)  in Euclidean space for such transformations can be 
evaluated for some 
particular choice  of the finite field dependent parameter, $\Theta[\phi(x)]$.
Such Jacobian contribution can be replaced (within the functional 
integral) by $ \exp[-S_1[\phi(x,\kappa) ]],$
 iff the following condition is satisfied \cite{sb}
\begin{equation}
\int D\phi (x) \;  \left [ \frac{1}{J}\frac{dJ}{d\kappa}+\frac
{d S_1[\phi (x,\kappa )]}{d\kappa}\right ]\exp{[-(S_{GZ}+S_1)]}=0, \label{mcond}
\end{equation}
where $ S_1[\phi ]$ is some local functional of fields.
The infinitesimal change in the $J(\kappa)$ can be calculated using the equation \cite{sdj}
\begin{equation}
\frac{1}{J}\frac{dJ}{d\kappa}=-\int d^4x\left [\Sigma_\phi\left\{\pm \delta_b \phi (x,\kappa )\frac{
\partial\Theta^\prime [\phi (x,\kappa )]}{\partial\phi (x,\kappa )}\right\}\right ],\label{jac}
\end{equation}
where $\pm$ sign refers to whether $\phi$ is a bosonic or a fermionic field.

The FFBRST transformation corresponding to the BRST transformation given in Eqs. (\ref{sym}) and
(\ref{syma}) is then written as
 \begin{eqnarray}
\delta_b A_\mu^a &=& -{\cal D}_\mu^{ab}c^b\ \Theta;\   \delta_b c^a =\frac
{1}{2}gf^{abc}c^bc^c
 \Theta;\  
 \delta_b \bar c^a = B^a \Theta;\  \delta_b B^a =0; 
 \nonumber\\
   \delta_b\omega_i^a &=&0;\   \delta_b\varphi_i^a  =  -\omega_i^a  \Theta;\
\delta_b\bar\omega_i^a  = \bar \varphi_i^a  \Theta;\   \delta_b\varphi_i^a
 = 0;\ 
 \delta_b U^{ai}_\mu  = M_\mu^{ai}  \Theta;\      
\nonumber\\  
\delta_b T^{ai}_\mu  &=& -R_\mu^{ai} \Theta;\ \delta_b V^{ai}_\mu  = -N_\mu^{ai} \Theta;\ \delta_b M^{ai}_\mu =\delta_b N^{ai}_\mu =  \delta_b R^{ai}_\mu =0, 
\end{eqnarray}
where $\Theta$ is finite, field dependent, anticommuting and explicit space-time independent parameter.
 
\section{Connecting GZ theory and YM theory in BV formalism}
The generating functional of YM theory in the BV formulation can be written by  introducing
antifields $\phi^\star $ corresponding to the all fields $\phi$
 with opposite statistics as,
{\begin{eqnarray}
Z_{YM} = \int [D\phi] e^{-\int d^4x\left\{\frac{1}{4}F^a_{\mu \nu }F_{\mu \nu}^a
+A_\mu^{a\star}{\cal D}_\mu^{ab}c^b 
+ \bar c^{a \star}B^a \right\}}.
\end{eqnarray}}
This can further be written in compact form as
 \begin{equation}
Z_{YM} = \int [D\phi] e^{-\left  [W_{\Psi_1 }(\phi,\phi^\star)\right]},
\end{equation} 
where the expression for gauge-fixing fermion  is  given as 
$\Psi_1 =  \int d^4x\ \left[\bar c^a\partial_\mu A_\mu^a \right] $.
The generating functional does not depend on the choice of gauge-fixing fermion \cite{ht}.
The extended quantum action, $W_{\Psi_1}(\phi,\phi^\star)$, satisfies certain rich mathematical
 relation called quantum master equation \cite{wei}, given by
\begin{equation}
\Delta e^{-W_{\Psi_1}[\phi, \phi^\star ]} =0  \ \mbox{ with }\ 
 \Delta\equiv \frac{\partial_r}{
\partial\phi}\frac{\partial_r}{\partial\phi^\star } (-1)^{\epsilon
+1}.
\label{mq}
\end{equation}
The antifields  can be evaluated from { $\Psi_1$} as
{ \begin{eqnarray}
A_\mu^{a\star }=\frac{\delta\Psi_1}{\delta A_\mu^a}= -\partial_\mu\bar c^a,\ \
 \bar c^{a\star}=\frac{\delta\Psi_1}{\delta \bar c^a}= 
 \partial_\mu A_\mu^a,\ \ B^{a\star}=\frac{\delta\Psi_1}{\delta B^a}= 
 0.
\end{eqnarray}}
Similarly, the generating functional of GZ theory in BV formulation can be written as, 
{\begin{eqnarray}
Z_{GZ} & =&  \int [D\phi] \exp\left[ -\int d^4x\left\{\frac{1}{4}F^a_{\mu \nu }F_{\mu \nu}^a
+A_\mu^{a\star}{\cal D}_\mu^{ab}c^b  
 +  \bar c^{a \star}B^a -\varphi_i^{b\star}\omega^b_i
 \right.\right.\nonumber\\
&+&\left.\left. +\bar\varphi_i^a\bar\omega_i^{a\star}+U_\mu^{ai\star}M_\mu^{ai} 
-V_\mu^{ai\star}N_\mu^{ai} 
 -   T_\mu^{ai\star}R_\mu^{ai} \right\}\right].
\end{eqnarray}}
This can further be written in compact form using gauge-fixed fermion ($\Psi_2$) as 
\begin{eqnarray}
Z_{GZ} &= &\int [D\phi] e^{-\left  [W_{\Psi_2 }(\phi,\phi^\star)\right]}, 
\Psi_2  = \int d^4x  \left[\bar c^a\partial_\mu A_\mu^a +
\bar \omega^a_i\partial_\mu {\cal D}_\mu^{ab}  
\varphi_i^b   \right.\nonumber\\ 
  &-&\left. U_\mu^{ai}{\cal D}_\mu^{ab}\varphi^{  bi} -V_\mu^{ai}{\cal D}_\mu^{ab}\bar\omega^{  bi}- U_\mu^{ai}V_\mu^{ai} 
+   gf^{abc}
T_\mu^{ai}{\cal D}^{bd}_\mu c^d\bar\omega^{ ci} \right]. 
\end{eqnarray}
The antifields are
obtained from  $\Psi_2$  as 
{ \begin{eqnarray}
A_\mu^{a\star } =  -\partial_\mu\bar c^a -gf^{abc}\partial_\mu\omega^{c i}\varphi^{ bi} 
-gf^{abc}U_\mu^{ci}\varphi^{  bi}   
-gf^{abc}V_\mu^{ci}\bar\omega^{  bi}, \ \
 \bar c^{a\star} =  
 \partial_\mu A_\mu^a,\nonumber \\ 
U_\mu^{ai
\star} =  -{\cal D}_\mu^{ab}\varphi^{  bi}  -V_\mu^{ai},\ \
\bar\omega_i^{a\star}= \partial_\mu {\cal D}_
\mu^{ab}\varphi^b_i -V_{\mu i}^{b }{\cal D}_\mu^{ba}
+ gf^{abc} 
T_{\mu i}^{b }{\cal D}^{cd}_\mu c^d,\nonumber\\ 
V_\mu^{ai\star}  =  -{\cal D}_\mu^{ab}\bar\omega^{  bi}
-U_\mu^{ai},\ T_\mu^{ai\star} =gf^{abc}
{\cal D}^{bd}_\mu c^d\bar\omega^{ ci},\  
\varphi_i^{b\star } =\bar \omega^a_i\partial_\mu {\cal D}_\mu^{ab} 
 -U_{\mu i}^{a}{\cal D}_\mu^{ab}.
\end{eqnarray} 
To connect these two theories we construct following finite field dependent parameter $\Theta [\phi]$ 
\begin{eqnarray}
\Theta [\phi] =\int_0^1 d\kappa \int d^4x \left[\varphi_i^{b\star}
\varphi_i^b +\bar\omega_i^{b\star}\bar \omega_i^{  b} 
  + V^{ai\star}_\mu V^{ai}_\mu \right].
\end{eqnarray} 
The Jacobian of path integral measure  in the generating functional (\ref{zfun}) for the FFBRST 
with this parameter can be replaced by $e^{-S_1}$ iff condition (\ref
{mcond}) is satisfied.
To find $S_1$ we start with an ansatz for $S_1$ as 
\begin{eqnarray}
S_1 = \int d^4x \left[\chi_1 \varphi_i^{b\star}\omega^b_i
+\chi_2 \bar\varphi_i^a\bar\omega_i^{a\star} + \chi_3  U_\mu^{ai\star}M_\mu^{ai} 
+\chi_4  V_\mu^{ai\star}N_\mu^{ai}
+\chi_5  T_\mu^{ai\star}R_\mu^{ai} 
\right].
\end{eqnarray}
where $\chi_j(\kappa) (j=1,2,..,5)$ are arbitrary but $\kappa$-dependent constants and satisfy 
following 
initial conditions 
$\chi_j(\kappa =0)=0.$
These constants are calculated using Eq. (\ref{mcond}) subjected to the initial condition to find the $S_1$ as
\begin{eqnarray}
S_1=\int d^4x \left[+ \kappa\ \varphi_i^{b\star}\omega^b_i
-\kappa\ \bar\varphi_i^a\bar\omega_i^{a\star}  -\kappa\ U_\mu^{ai\star}M_\mu^{ai} 
+ \kappa\  V_\mu^{ai\star}N_\mu^{ai}
+ \kappa\  T_\mu^{ai\star}R_\mu^{ai} 
\right].
\end{eqnarray}
By adding $S_1(\kappa=1)$ to $S_{GZ}$, we get
$
S_{GZ}+S_1(\kappa=1)= S_{YM}.
$
Hence, 
\begin{eqnarray}
Z_{GZ}\left(\equiv \int [D\phi]\ e^{-W_{\Psi_2 }} \right)  \stackrel{FFBRST}{----\longrightarrow }
Z_{YM}\left(\equiv \int [D\phi]\ e^{- W_{\Psi_1 }(\phi,\phi^\star) }\right) 
\end{eqnarray}
Thus, using  FFBRST transformation we connect the GZ theory to YM theory in BV formulation. 
\section{Conclusion}
  The KO criterion for color 
confinement in a manifestly covariant gauge is not satisfied in GZ theory due to  the presence of the horizon term
which breaks the usual BRST symmetry. However this theory is extended to restore the BRST symmetry and hence the 
KO analysis is satisfied for color confinement. This BRST symmetry is generalized by
allowing the transformation parameter finite and field dependent in the context of BV formulation.
 Such FFBRST 
transformation is also symmetry of the effective action $S_{GZ}$ but not of the generating functional $Z_{GZ}$.
We have shown that the FFBRST with an appropriate choice of finite field dependent parameter relates GZ theory 
with a correct horizon term 
to the YM theory in the BV formulation. Thus  the 
theory free from Gribov copies has shown to be related 
 to a theory with Gribov copies (i.e. YM 
theory) through FFBRST transformation in the BV formulation. 
The nontrivial Jacobian of path integral measure in  $Z_{GZ}$ is responsible for this important 
connection.\\

{\Large{\bf {Acknowledgment}}}

{One of us (SU)  acknowledge the financial assistance from  DST, India under ITS scheme (grant No. 
R/ITS/3443/2011-2012) and organizers of ``$8^{\mbox{th}}$ International Conference on Progress in Theoretical 
Physics" Constantine, Algeria.} 

\end{document}